\documentclass[a4paper,american]{scrartcl}

\usepackage{tikz-cd}
\usepackage{verbatim}
  \usepackage{enumerate}
\usepackage{lmodern}
\usepackage[utf8]{inputenc}
\usepackage[T1]{fontenc}
\usepackage{color}
\usepackage{todonotes}
\usepackage{babel}
\usepackage{amsfonts, amsmath, amsthm, amssymb}
\usepackage{setspace}
\usepackage[authoryear]{natbib}
\usepackage[font=small]{quoting}
\usepackage{latexsym}
\usepackage[bookmarks=false,pdfborder={0 0 0},colorlinks=false]{hyperref}
  \usepackage{dsfont}
  \usepackage{xcolor}

  \makeatletter
  
  \usepackage[arrow, matrix, curve]{xy}

  \newcommand{\IR}{\mathbb R}

    \newcommand{\Div}{\mathrm{div}}

  \theoremstyle{plain}

  \theoremstyle{definition}

\title{Against Fields} 
\author{Dustin Lazarovici \\ Université de Lausanne, Faculté des Lettres\\
	Section de Philosophie, 1015 Lausanne, Switzerland\\
	Dustin.Lazarovici@unil.ch}
\begin{document}

\maketitle
\begin{abstract} 
\noindent Using the example of classical electrodynamics, I argue that the concept of fields as mediators of particle interactions is fundamentally flawed and reflects a misguided attempt to retrieve Newtonian concepts in relativistic theories. This leads to various physical and metaphysical problems that are discussed in detail. In particular, I emphasize that physics has not found a satisfying solution to the self-interaction problem in the context of the classical field theory. To demonstrate the superiority of a pure particle ontology, I defend the direct interaction theory of Wheeler and Feynman against recent criticism and argue that it provides the most cogent formulation of classical electrodynamics. 
\end{abstract}
\section{Introduction}

The story of field theory -- and modern physics is, to a large extent, field theory -- is often told as a tale of triumph. The field concept (it seems) is one of the most successful in all of science. It has shaped our understanding of symmetries and interactions. It has survived, if not driven, all major physical revolutions of the past century and is now a central part of our most thoroughly tested theories. 

The aim of this paper is to put forward a different narrative: while fields have proven to be extremely successful \emph{effective} devices, physics has time and time again run into an impasse by promoting these devices to a fundamental level and buying into a dualistic ontology of fields and particles. On closer examination, the concept of fields as mediators of particle interactions turns out to be philosophically unsatisfying and physically problematic, as it leads, in particular, to problematic self-interactions. Against this background, I will argue that the true significance of fields is that of ``book-keeping variables'' (\cite{Feynman:1966aa}), summarizing the effects of retarded and/or advanced relativistic interactions in order to provide an efficient description of subsystems in terms of \emph{instantaneous sates}. And while this works very well for practical purposes, it is ultimately at odds with the principles of relativity.

Since this thesis is in stark opposition to the current mainstream view in physics and philosophy, a careful analysis will be necessary to defend it. I will focus on Maxwell-Lorentz electrodynamics as the \emph{locus classicus} of relativistic field theory, though most arguments apply \emph{mutatis mutandis} to other field theories, including quantum field theory, where the duality of particles and fields is mirrored by the duality of fermionic and bosonic ``fields'' and where the self-interaction problem persists in form of the infamous \emph{ultraviolet divergence}. However, since the ontology of QFT is a highly contentious subject in itself, I will avoid it for the rest of the paper. A complementary discussion might also be necessary for pure (monolithic) field theories, though I'm not sure if a serious contender is even on the table. 

The paper is structured as follows: Sections 2 and 3 set out the main arguments against fields. These can be divided into physical and metaphysical problems, the crucial difference being that metaphysical problems can be solved, or at least mitigated, by a mere \emph{reinterpretation} of existing field theories, while the solution of the physical problems requires new physics. Section 4 will revisit the most common arguments for fields and show that they remain inconclusive. Section 5 and 6 will explore the relationship between the field concept and relativistic space-time and discuss in what sense the electromagnetic field can be reduced to the history of charged matter. It is here that we find first indications, within Maxwell's theory itself, that fields are best conceived as an effective concept rather than a fundamental one. Section 7 will discuss the Wheeler-Feynman theory that abandons fields in favor of direct particle interactions. I will argue that this theory provides the most satisfying and coherent formulation of classical electrodynamics. Finally, by comparing the situation in Wheeler-Feynman and Maxwell-Lorentz theory, section 8 will highlight the role of fields in the formulation of \emph{initial value problems}, which I believe to be the real reason why physics is so attached to the field concept. However, as I will present recent mathematical evidence that the initial value paradigm fails in relativistic space-time, the field concept fails, as well.   

A forerunner of the present work is \cite{Mundy:1989aa}, who advocates an interpretation of classical electrodynamics in terms of direct interactions. While I agree with most of Mundy's conclusions, I believe that an even stronger case can be made by examining the field theory and its relationship to the direct interaction competitor more carefully.

\section{Against fields: The philosophical case}

Every precise physical theory has to spell out a clear ontology, that is, it has to provide a precise answer to the question: what is the theory about? Newton formulated his law of universal gravitation as a theory of point particles (``solid, massy, hard, impenetrable, moveable particles''). While it is possible to introduce a ``gravitational field'' (see section \ref{sec:eliminatefields}), this object appears only as a book-keeper of particle interactions rather than a candidate for ontology. When it comes to electrodynamics, the situation seems to be different. The electromagnetic field follows a dynamical law of its own (the Maxwell equations) and so classical electrodynamics seems to be as much a theory \emph{about} the electromagnetic field as it is a theory about charged particles interacting with and through the electromagnetic field.

However, the recent philosophical debate about the status of the wave function in quantum mechanics has shown how problematic it can be to read off the ontology of a theory from its formalism, as this approach often fails to distinguish sensible ontological commitments from mathematical representations of the laws (see e.g. \cite{ Maudlin2015}, p. 357). A different strategy has been worked out as the \emph{primitive ontology} approach (the term goes back to \cite[ch.2]{Durr:2013aa}, originally published 1992, cf. also Bell's notion of ``local beables'' in \cite[ch.7]{Bell:2004aa}, originally published in 1975). It starts from the observation that all empirical data (including, in particular, experimental data manifested in pointer positions, display readings etc.) can be understood as consisting of particle positions and particle motions, or, more generally, of the distribution of matter through space and time. This is to say that the empirical content of a theory is exhausted by its statements about the primitive ontology -- the fundamental entities posited as the basic constituents of matter. All other objects appearing in the formalism can then be understood as part of the \emph{dynamical} or \emph{nomological} structure, introduced to explain (or guide or describe) the evolution of the primitive variables, e.g., particle positions.

This strategy has been successfully applied to the nomological interpretation of the wave function in (Bohmian) quantum mechanics, leading, in particular, to the development of Quantum Humeanism (see e.g. \cite{Durr:2013aa}, ch.12, \cite{Miller:2013}, \cite{Esfeld:2014aa},  \cite{Esfeld:2014ab}). It is, however, neither committed to a particular theoretical framework nor a particular metaphysics of laws. Applied to classical electrodynamics, the primitive ontology approach leads to an interpretation of the theory that admits only the particles as physical entities and puts the fields on the side of the laws. The key argument for this interpretation is its sparse ontology that should appeal to everyone with the proverbial taste for desert landscapes. Subsequently, the usual proposals -- Humeanism, primitivsim and dispositionalism -- are on the table to ground the laws in the ontology. For instance, endorsing the supervenience thesis of Lewis (e.g. (1986) pp. ix--xi), one can conceive the Humean mosaic as being made up of particle positions and particle motions only, while the fields appear as part of the \emph{best system} that strikes an optimal balance between being simple and being informative in summarizing the mosaic.

The conservative position that admits an ontology of particles \emph{and} fields comes with no gain in explanatory power but faces the additional challenge to explain what exactly the electromagnetic field is supposed to be. Is it a property of space-time points, as proposed by \cite{Field:1985aa}? Is it some sort of stuff filling space-time?  

Before assessing any of these proposals, one should be clear about what exactly they are committing to. Quite often, philosophical discussions about fields refer implicitly or explicitly to \emph{scalar fields}. A scalar field is represented by a real-valued function, meaning that it is completely described by its ``strength'' at every point in space-time. This seems intuitive enough. If the field is stuff, then there is a little more stuff over here and a little less stuff over there. If the field is a property of space-time, then the property is instantiated with different magnitudes at different locations.

The electromagnetic field, however, is not a scalar field. It is not even a vector-field that would be characterized by a magnitude and a direction at every point in space. Electric and magnetic fields can be separately represented as vector-fields, but the splitting of the electromagnetic field into electric and magnetic components is possible only with respect to a particular frame of reference. Since electrodynamics is a relativistic theory, it turns out that the electromagnetic field is most accurately described by an anti-symmetric $4\times4$-tensor $F^{\mu \nu}$ that comprises electric and magnetic components and transforms canonically under Lorentz transformations. It is quite difficult to attach a physical interpretation to this abstract mathematical object other than through its role in determining the particle dynamics by figuring in the Lorentz force equation
\begin{equation}\label{Lorentzforce} m \ddot{z}^\mu = e\,  F^{\mu\nu} \, \dot{z}_\nu. \end{equation}
(Here, $m$ denotes the mass and $e$ the charge of the particle with space-time trajectory $z^\mu(\tau)$ and a dot denotes a derivative with respect to eigentime. We use natural units in which the speed of light is equal to one.)

However, if we understand the electromagnetic field only in terms of its effect on particles -- if we agree, in other words, that the field is introduced into the theory to explain the motions of particles -- it seems quite excessive as a physical / ontological structure: the field stretches out all over space-time, far into regions that contain no particles at all, and thus involves an infinite number of degrees of freedom that will never manifest as influences on particle motions. (You might say it is still useful for grounding counterfactuals, but is this not precisely the role of \emph{nomological} structures?)

Moreover, the two proposals ``fields as properties of space-time'' and ``fields as primitive stuff'' each face additional difficulties of their own. Geometrical properties are \emph{bona fide} properties of space-time, but space-time geometry, unlike the electromagnetic field, should concern the motions of \emph{all} particles. After all, it is precisely the universality of gravitation that motivates its geometric account in general relativity. Since the electromagnetic field interacts selectively, with charged particles only, the conception of fields as properties of space-time points entails the commitment to an exceptionally strong form of substantivalism that endows space-time with causal properties \emph{over and above} the geometrical ones. 

Also, if one conceives the field properties as dispositions, one faces the following problem: there is a dispositional property of the particles -- their charge -- which produces in the first place a further disposition -- the field properties -- and both these dispositions then manifest themselves in the acceleration of particles according to \eqref{Lorentzforce}. Consequently, every interaction mediated by a field involves two distinct dispositional properties that bring about the same manifestation.

If, on the other hand, one regards the field as another substance, as some sort of \emph{primitive stuff}, the nature of this stuff remains obscure. In the end, the substance view comes down to positing primitive field stuff as a \emph{bare substratum} that holds together the (dispositional) field properties. This does not elucidate the meaning of the field beyond its nomological role but leads to unnecessary problems like the question, how to interpret the field value zero. Is there \emph{no} stuff in space-time regions where $F_{\mu \nu} =0$? Or does the field stuff exist everywhere and merely exerts no force in those regions?

In conclusion, the conservative options make additional ontological commitments that do not come with any gain in explanatory power but create new drawbacks instead. Hence, I maintain that the clearest and most coherent interpretation of electrodynamics is one in which the electromagnetic field, strictly speaking, does not exist (or only in the sense in which other structures, figuring in the particle laws of motion, exist). Of course, if the duality of particles and fields is philosophically unsatisfying, labeling the field as a nomological or dynamical structure is merely a spoonful of sugar that helps the medicine go down. A more consequential solution would require a reformulation of the theory in which the field does not appear in the first place. As I will argue in the course of this paper, such a reformulation of classical electrodynamics is suggested not only by philosophical considerations but, more importantly and more emphatically, by physical ones.

 \section{Against fields: The physical case}\label{sec:physics}

The Maxwell-Lorentz theory of classical electrodynamics consists of two parts. The Maxwell equations describe the evolution of the electromagnetic field and its coupling to charges and currents. The Lorentz force equation \eqref{Lorentzforce} describes the motion of a charge in the presence of an electromagnetic field. The field equations can be further separated into homogeneous and inhomogeneous equations. The homogeneous equations tell us that the field tensor $F^{\mu \nu}$ (a 2-form) can be written as the exterior derivative of a vector-potential (a 1-form) $A^\mu$, i.e. as
\begin{equation}\label{field tensor} F^{\mu\nu} = \partial^{\mu} A^{\nu} - \partial^{\nu} A^{\mu}. \end{equation}
The inhomogeneous Maxwell equations tell us how the electromagnetic field is influenced by charges. Fixing the gauge-freedom in \eqref{field tensor} by demanding $\partial_\mu A^\mu(x) = 0$ (Lorentz gauge), they take the particularly simple form:
\begin{equation}\label{Maxwell} \square A^{\mu} = 4 \pi j^{\mu}, \end{equation}
\noindent with $\square = \partial_\mu\partial^\mu$ the d'Alembert operator and $j^{\mu}$ the 4-current density, which for $N$ point charges is:
\begin{equation}\label{current} j^\mu(x) = \sum_{i=1}^{N} j^\mu_i(x) =  \sum\limits_{i=1}^{N}e_i \int \delta^4(x-z_i(\tau_i))\dot{z}_i^\mu(\tau_i) \, \mathrm{d}\tau_i. \end{equation}

\noindent \textit{Given} the charge trajectories $z_i(\tau_i),\; i=1,...,N$, the solutions of \eqref{Maxwell} are well known. By linearity of \eqref{Maxwell}, we can sum the contribution of each particle. A special solution is given by the advanced and retarded \textit{Li\'enard-Wiechert} potentials
\begin{equation}\label{LW} A^\mu_{i,\pm}(x) =   \frac{e_i \, \dot{z}_i^\mu(\tau^\pm_i)}{\bigl(x^\nu - z_i^\nu(\tau_i^\pm)\bigr) \dot{z}_{i,\nu}(\tau_i^{\pm})},\end{equation}
where $x \in \IR^4$ denotes a space-time point and $\tau^+_i(x)$ and $\tau_i^-(x)$ are the \emph{advanced} and \emph{retarded} times given as implicit solutions of
\begin{equation}\label{taupm} \bigl(x^\mu-z^\mu_i(\tau)\bigr)(x_\mu-z_{i,\mu}(\tau)\bigr) = 0. \end{equation}
\noindent This means that the field equations connect events with Minkowski distance zero, so that the advanced / retarded field at $x$ depends on the charge trajectories at their points of intersection with the future, respectively the past light cone of $x$. 

Now any solution of the field equations can be written as a convex combination of advanced and retarded Li\'enard-Wiechert field plus a solution of the free wave equation
\begin{equation}\label{Maxwellfree} \square A^{\mu} = 0, \end{equation}
corresponding to a so-called free field. 

A self-consistent description, however, requires us to solve \eqref{Maxwell} and \eqref{Lorentzforce} \textit{together}. And this set of coupled equations is ill-defined. As we can see from \eqref{LW}, the electromagnetic field is singular precisely at the points where it has to be evaluated in \eqref{Lorentzforce}, that is, on the world-lines of the particles. This is the notorious problem of the \textit{electron self-interaction}: a charged particle generates a field, the field acts back on the particle, and since the field-strength is infinite at the position of the particle, the interaction blows up. Note that even the one-body problem is ill-defined. It is not the interaction between particles but the duality of particle and field that leads to singularities.

The reason why classical electrodynamics still works so well for practical purposes is that physicists, in general, solve the Maxwell equations for a given charge distribution or the Lorentz equation for a given electromagnetic field but not both together in a self-consistent way (cf. \cite{Frisch2004}). Strictly speaking, Maxwell-Lorentz electrodynamics is an inconsistent theory.

Before discussing possible ways of dealing with the self-interaction problem, another physical issue should not go unnoticed. Electromagnetic radiation, propagating at the speed of light, will eventually overtake all particles and thus carry energy away to infinity. For this reason, a Maxwell-Lorentz universe is in some sense like a dissipative system and does not allow any stable bound states (\cite{KomechSpohn}, see in contrast \cite{Schild} for the existence of bound states in the Wheeler-Feynman direct interaction theory). This gives further support to the central theme of this paper: that the field degrees of freedom are phony and physically problematic.

\subsection{Solving the self-interaction problem}

Throughout the 20th century, various attempts have been made to solve the self-interaction problem in order to obtain a well-defined theory of the classical electron. Let's take a look at the most important ones.\\

\noindent \textbf{Maxwell-Lorentz without self-interaction.} The version of classical electrodynamics that is (implicitly) used in most practical applications can be described as Maxwell-Lorentz without self-interaction (ML-SI): when evaluating the Lorentz-force law \eqref{Lorentzforce}, a charge is taken to interact only with the electromagnetic fields produced by \emph{other} charges, ignoring its own contribution in the field equation \eqref{inhomo}. The equations of motion for an $N$-particle system thus read: 
\begin{equation}\label{ML-SI} 
 m_k \ddot{z}^\mu_k =  \sum\limits_{j \neq k} e_k e_j\, {}^{(j)}F^{\mu\nu} \dot{z}_{k,\nu}, \end{equation} 
where $ {}^{(j)}F$ denotes the field attributed to particle $j$ and the self-field is notably absent in the sum on the right-hand-side. While this pragmatic solution works well for most practical purposes (in particular on macroscopic scales) there are at least two problems that prevent us from taking it seriously on a more fundamental level. 

\begin{enumerate}[1)]

\item ML-SI does not account for the damping force experienced by an accelerated charge. This so-called \emph{radiation damping}, that is necessary for the energy balance of a radiating particle, is usually attributed to the back-reaction of the field and is thus absent in a description that neglects such effects altogether.\footnote{This seems to be the basis of the inconsistency claim in \cite{Frisch2004}.}

\item As we can see seen from equation \eqref{ML-SI}, ML-SI involves not just \emph{one} electromagnetic field but a different field for every single particle in the universe. The field created by particle $j$ must carry some property that makes it interact with all the other charges but not with $j$ itself. This leads to an enormous increase in mathematical complexity of the $N$-body problem (we have to specify $N$ initial fields and track the evolution of each one separately) as well as to a grotesque inflation of the physical ontology. 

\end{enumerate}

\noindent \textbf{Extended particle models.} Early attempts modeled the electron as an extended charge distribution rather than a point-particle. In particular, there is the semi-relativistic Abraham model (which assumes a rigid charge distribution in some preferred frame or, alternatively, in all frames, ignoring the relativistic Lorentz-contraction) and the more sophisticated Lorentz model, which assumes a rigid charge distribution in the momentary rest-frame of the particle (\cite{Abraham}, \cite{Lorentz2}, see \cite{Spohn} for a state-of-the-art discussion). While these theories are free of singularities, they are rarely taken seriously for the following reasons: 

\begin{enumerate}[1)]

\item There is no empirical evidence for an internal structure of the electron that would justify a departure from the point-particle model. (Experiments put the upper bound in a possible electron radius to $10^{-22}m$, see \cite{Dehmelt}.) 

\item Extended charges models introduce effects depending on size and shape of the particles that seem ad hoc and unwarranted in the absence of empirical evidence.

\item The models are only consistent with the relativistic energy-momentum relations if one postulates internal forces of yet unknown origin (``Poincar\'e stresses'') that hold the charge distribution together (see e.g. \cite{Feynman1963}, ch. 28). 

\item For large accelerations, a fully relativistic treatment of extended charges leads to acausal artifacts like parts of the charge distribution overlapping with itself and appearing to move backwards in time in certain Lorentz frames (\cite{Nodvik}, \cite{Rohrlich2007}, ch. 7-4).
\end{enumerate}

\noindent \textbf{Lorentz-Dirac theory.} The equation that is believed to capture all classical radiative phenomena is the Lorentz-Dirac equation
\begin{equation}\label{Lorentz-Dirac} m \ddot{z}^\mu = e F^{\mu \nu} \dot{z}_\nu + \frac{2}{3}e^2 (\dot{z}^\mu \ddot{z}^\nu\ddot{z}_\nu- \dddot{z}^\mu ), \end{equation}
	\noindent where $F^{\mu \nu}$ on the right-hand-side does not contain the self-field and the 4-vector 
	\begin{equation}
	\label{Schott}
\Gamma^\mu := \frac{2}{3}e^2 (\dot{z}^\mu \ddot{z}^\nu\ddot{z}_\nu-\dddot{z}^\mu )	\end{equation}
includes, in particular, the radiation damping. 

In his seminal paper on the classical electron, \cite{Dirac} showed that \eqref{Lorentz-Dirac} can be derived from Maxwell's equations together with the principle of energy-momentum conservation; analogous results can be obtained by considering the point-particle limit for spherical charge distributions (see \cite{Rohrlich1997} for a good historical overview). These derivations, however, rely on a highly dubious \emph{mass renormalization} procedure: an infinite, negative bare mass has to be introduced to cancel a diverging inertial term arising from the self-energy. Moreover, while the derivations suggest the self-interaction as the origin of the radiation reaction, this is not a consistent interpretation of the final theory. The right-hand-side of \eqref{Lorentz-Dirac} is divergence-free at the position of the particle and hence does not contain any self-field according to Maxwell's equations (cf. \cite{Wheeler-Feynman:1945aa}, p. 159). 

That said, we would not have to take the derivation seriously in order to accept \eqref{Lorentz-Dirac} as the fundamental law of motion for a classical point charge. Unfortunately, the Lorentz-Dirac theory still has several grave issues (cf. \cite{Frisch2005}, ch. 3.3): 

\begin{enumerate}[1)]
	\item Except for very fine-tuned initial conditions, the triple-dot term in \eqref{Lorentz-Dirac} leads to \emph{runaway solutions} that approach the speed of light exponentially fast and run off to infinity. This can be understood as a manifestation of the infinite self-energy in form of perpetual \emph{self-acceleration}. 
	
	Note that the issue is not just that the Lorentz-Dirac equation has clearly unphysical solutions (so does Newtonian gravity) but that these solutions are in some sense \emph{typical}, indicating that the mass renormalization program is fundamentally flawed. 
	
	\item Another pathology of the Lorentz-Dirac equation are \emph{pre-accelerations}, meaning that particles can start to accelerate before being subject to exterior forces (see \cite{Grunbaum1976} for a good discussion). While not as problematic as runaway behavior, pre-accelerations are certainly counterintuitive and violate Newton's law of inertia. 

	\item Accepting \eqref{Lorentz-Dirac} as the new particle law of motion, we still have to give up the superposition principle for fields in order to exclude the self-field. The Lorentz-Dirac theory thus shares with ML-SI the unbecoming feature that it involves not one electromagnetic field but a different field associated to each and every particle. 
	
\end{enumerate}

\noindent \textbf{Non-linear field theories.} While the previous proposals remained within the framework of Maxwell's theory, it is possible to follow a more radical approach and replace the Maxwell equations with non-linear field equations such as those proposed by \cite{BornInfeld}. The Born-Infeld equations are Lorentz invariant, have finite self-energy solutions for point charges and approximate the Maxwell system in the weak field limit. However, the corresponding Lorentz-force is still ill-defined, so that alternative laws of motion have to be investigated (see \cite{Kiessling2012} for a recent discussion). Moreover, the non-linear field equations are extremely difficult to handle and thus not very well studied. 

I mention Born-Infeld here to acknowledge the fact that it is not \emph{a priori} true that field theories lead to self-interaction divergences. How far the merits of the theory extend beyond that is unclear and in need of further physical and mathematical investigation. In any case, the Born-Infeld theory is not what people usually have in mind when they advocate the successes of field theories.\\ 

\noindent In conclusion, in one and a half centuries, physics has not produced a satisfying theory of interacting particles within the perimeters of the classical field concept.\footnote{I would strongly contest the common believe that the situation is much better in quantum theory, however not here.} Of course, we did not discuss the various (often mutually incompatible) proposals for dealing with the problems of self-acceleration and pre-acceleration encountered in the Lorentz-Dirac theory, some of which are significant technical achievements (see in particular \cite{Muller} for an overview of results and a staunch defense of the classical field theory). But when all is said and done, the conclusion remains that there is a fundamental conflict between the particle and the field concept that can be doctored at various levels of rigor and sophistication yet not completely resolved (\cite{Muller}, pp. 267--268 seems to agree). To my mind, a cogent physical theory is one that just needs to be \emph{analyzed} -- not \emph{fixed} along the way. As we will see in section \ref{sec:WF}, such a theory of classical electrodynamics actually exists. The key is to abandon the source of all evil -- the electromagnetic field -- in favor of direct particle interactions.

\section{Arguments for fields}
Having laid out the case against fields, let us revisit the most common arguments presented in their favor. Since the respective debates are well documented in the literature, I will comment on them only briefly and refer the reader to \cite{lange2002introduction} and \cite{Pietsch} for good and thorough discussions. 

\subsection{Light obviously exists}
The proposal of a pure particle ontology runs counter to the intuition that the existence of (at least some) electromagnetic fields is somehow ``obvious''. After all, electromagnetic fields are \emph{obviously} there when we turn on the radio. More importantly, there seems to be a quite literal sense in which all we actually \emph{see} is light. And light can \emph{obviously} be manipulated: it can be reflected, refracted, polarized, absorbed.... Against this background, the claim that light does not exist because the electromagnetic field does not exist may seem absurd. 

The program defended in this paper is, however, not incompatible with the use of a language that contains \emph{light}, \emph{radiation}, \emph{electromagnetic signals}, etc. It rather insists that, in the language of our fundamental theory, propositions involving these terms should reduce to propositions about interactions between material objects, ultimately particles. To put it differently, such propositions remain true but their truth-maker are particle motions only.

The ``electromagnetic signal'' that we pick up when we turn on the radio refers to a particular kind of interaction between the receiver and coherently oscillating particles at the broadcasting station. An observation of ``red light'' refers to a particular kind of interaction between the observed object and our visual receptors. The ``reflection of light'' refers to a series of interactions involving a source $S$ a mirror $M$ and a target $T$ so that certain counterfactuals about the strength of the effect on $T$ depending on the presence/absence of $M$ and the geometry of the setup are true. It is somewhat tedious but straightforward to spell out the details.

As far as the basic phenomenology of light is concerned, science has already pushed us down this eliminative path. Our physical theories do not contain any \emph{redness} or \emph{blueness} or \emph{greenness}, instead, sense data involving the perception of color are related to specific frequencies in the electromagnetic field. To eliminate the electromagnetic field, as well, in favor of a theory of interacting particles now merely means to cut out the middle-man: the frequencies that we identify with red or blue or green light are taken to refer directly to particle accelerations.

Adopting the terminology of \cite{Sellars} we can say: our \emph{manifest image} and our \emph{scientific image} of the world connect at different points than they used to with the field theory, but this doesn't make the manifest image less accurate or the theoretical account less compelling. 

\subsection{Fields carry conserved quantities} A popular line of argument suggests that fields are indispensable because they ensure the conservation of energy-momentum in relativistic theories. In brief, the reasoning is that since relativistic interactions between particles are delayed, the potential energy must be stored \emph{somewhere}, and this \emph{somewhere} is the (e.g. electromagnetic) field.

The argument has, in fact, a physical and a philosophical part. The physical part is: fields are necessary to obtain conservation laws. The philosophical part is: since fields carry conserved quantities, they must be \emph{real}. Both conclusions are premature.  

A ``conserved quantity'' is in the first place a mathematical expression, a \emph{first integral} of the equations of motion. Its relevance lies in the fact that it partitions the solution space of the theory, thus constraining the possible motions of particles. There is, however, nothing about a conservation law that compels us to interpret the conserved quantity -- or the mathematical objects appearing in its definition -- as real physical entities that exist over and above the particles. 

That said, due to the self-energy problem, the field theory of classical electrodynamics does not even ensure the conservation of energy-momentum in any meaningful sense, as witnessed by the possibility of runaway behavior. On the other hand, the direct interaction theory of Wheeler and Feynman does support a notion of energy-momentum conservation without containing any fields at all (\cite{Wheeler-Feynman:1949}). The respective quantities supervene on larger segments of trajectories rather than instantaneous states (cf. section \ref{sec:IVP}) but serve very much the same purpose as their Newtonian counterparts. 

In conclusion, fields are neither necessary nor sufficient for energy-momentum conservation, and even when they do figure in conservation laws, nothing of philosophical interest follows from that.

\subsection{No action at a distance}

The case for fields is often cast as a case against \emph{action at a distance}. The concept of particles interacting directly, without mediation by another substance, is then usually rejected as mysterious or absurd (even if the distant actions are \emph{delayed} as in the context of electrodynamics). However, as \cite{lange2002introduction}, pp. 94--95 noted, actual arguments showing the alleged absurdity are remarkably difficult to find, except for circular reasonings along the lines of ``action at a distance is unacceptable, because something cannot act where it is not present''. 

Historically, the development of field theory was deeply rooted in the mechanistic thinking of the early 19th century. However, the respective intuitions did not pass the test of time and with the demise of the ether theory, it is doubtful whether the modern field concept even addresses the worries regarding action at a distance that were voiced by \cite{Maxwell} and contemporaries. Upon the most reasonable reading, the argument that still resonates today maintains that an action at a distance theory has to take the causal / dynamical relations between the particles as primitive, while the field theory offers a deeper, or at least more detailed explanation, telling us \emph{how} particles affect each other. But this argument is circular, as well, since it presupposes that local interactions are more explanatory than distant ones. 

From a philosophical point of view, one could turn the table and make the case that dynamical relations between two tokens of the same type are more explanatory than interactions between different kinds of substances, viz particles and fields. Moreover, the field theory has to take \emph{two} kinds of interactions as primitive: charges generating fields and fields accelerating charges.

In any case, our best metaphysics of laws today have no difficulties accommodating action at a distance, i.e., fundamental laws connecting particle events with non-zero spatial or spatio-temporal distance. The primitivist takes action at a distance as primitive, while the Humean can adopt it as part of the best system. The dispositionalist, who seeks to ground the laws in causal properties of the particles, may find action at a distance somewhat mysterious if he construes the dispositions as \emph{intrinsic} properties. But this is a pseudo-problem that disappears if the dispositions are understood, more aptly, as dynamical relations (see \cite{Esfeld2009}). 

In summary: There is no problem with action at a distance and the field theory does not solve it.

\section{Fields and relativistic space-time}\label{sec:eliminatefields}

\noindent The history of modern field theory is intimately linked to the rise of relativity. After all, Maxwell's theory was relativistic before we even knew what it meant to be relativistic, that is, before Einstein had the audacity to propose that the symmetries discovered in Maxwell's equations reflect a fundamentally new structure of space and time. Nonetheless, one of the central goals of this paper is to convince the reader that the field concept ultimately reflects a failure to take the implications of relativity seriously enough.

To appreciate the  connection between the field concept and relativistic space-time, it is instructive to compare the electromagnetic field with the ``gravitational field'' in Newtonian mechanics. I put ``gravitational field'' in quotation marks, because Newton's theory of gravity is not a field theory in the modern sense but rather an action at a distance theory on classical space-time. Nevertheless, it is often convenient to introduce the gravitational potential 
\begin{equation}\label{potential} V(x) = - \sum\limits_{i=1}^N \frac{G m_i}{\lvert x - x_i \rvert}, \end{equation}
respectively the corresponding force field
 \begin{equation}\label{forcefield}
 -\nabla_x V(x) = - \sum\limits_{i=1}^N \frac{G m_i (x - x_i)}{\lvert x - x_i \rvert^3}.
\end{equation}

\noindent In the physical literature, the meaning of these fields is usually explicated in terms of a hypothetical \emph{test particle} that experiences a gravitational force from the $N$ massive particles \emph{without gravitating itself}. The necessity of this, quite unphysical, hypothetical is also due to the fact that the value of \eqref{forcefield} is infinite at any point $x_1, ..., x_N$ occupied by a particle that actually takes part in the interactions. The reason why these singularities are not really problematic, is that the gravitational field does not enter the Newtonian laws of motion as independent degrees of freedom. While it can be a useful mathematical tool, the gravitational field is just a book-keeper of direct particle interactions rather than a physical entity that exists over and above the particles.

To see how far the analogy between gravitational and electromagnetic fields goes, we recall that the potential \eqref{potential} can be derived as a solution of the \emph{Laplace equation}, which is arguably the simplest partial differential equation invariant under Euclidean symmetries. That is, outside the particle locations, the Newtonian potential satisfies  
 \begin{equation}
 \Delta V(x) = 0,
 \end{equation}
 while in the presence of point-particles at $x_1, ...,x_N$,
 \begin{equation}\label{Laplace}
 \Delta V(x) = 4 \pi G \sum \limits_{i=1}^N m_i \delta(x-x_i),
 \end{equation}
 with $\delta$ the Dirac delta-function, indicating that the masses are point-sources of the gravitational field. 
 
 On relativistic space-time, the Laplacian $\Delta = \delta^{ij} \partial_i \partial _j$ has a natural counterpart in the d'Alembert operator $\square = \eta^{\mu \nu} \partial_\mu \partial_\nu$, where $\eta_{\mu \nu}$ is the Minkowski metric and we adopt the convention of summing over double indices. This gives rise to what is arguably the simplest Lorentz-invariant PDE, namely, in empty space,
\begin{equation}\label{homo}
\square A^\mu (x) = 0,
\end{equation}
where $x$ is now a four-dimensional space-time variable, while along the charge trajectories
\begin{equation}\label{inhomo}
\square A^\mu (x) = 4 \pi j^\mu(x),
\end{equation}
with $j^\mu$ as in \eqref{current}, indicating that 4-currents are the sources of the, now electromagnetic, field. Of course, \eqref{inhomo} is precisely the Maxwell equation \eqref{Maxwell} which can thus be understood as the natural relativistic generalization of Laplace's equation.

Given this formal analogy, one must wonder why the pertinent theories treat the solutions of \eqref{Laplace} and \eqref{inhomo} (respectively their exterior derivatives) on such a different footing. Why, in other words, do we attribute to the electromagnetic field an ontological status and a causal role that is evidently unjustified in case of the gravitational field? There are two important facts that explain (but not justify) this difference:

\begin{enumerate}
	
\item[(1)] The Laplace equation \eqref{Laplace} is such that the sources influence the field \emph{instantaneously}, that is, along equal time hypersurfaces, while the relativistic equation \eqref{inhomo} is such that sources affect the field in a \emph{retarded} and / or \emph{advanced} way, that is, along past and / or future light cones. 

\item[(2)] Excluding unphysical potentials that diverge at infinity, \eqref{potential} is the \emph{unique} solution of \eqref{Laplace}, modulo an additive constant that does not affect the induced gravitational force and thus appears as a mere gauge symmetry. The Maxwell equation, in contrast, admits a variety of non-trivial and non-equivalent \emph{vacuum solutions} (solutions of \eqref{homo}) that can be added to any particular solution of \eqref{inhomo}. Such \emph{free fields} do affect the total electromagnetic force and are thus causally efficacious. 
\end{enumerate}

\noindent I want to stress that (1) is, in fact, exactly what one should expect if one takes the space-time geometry seriously. Galilean laws connect events on equal time hyperplanes, drawing on the structure of Newtonian or Galilean space-time, while relativistic laws connect events along light cones, drawing on the structure of Minkowski space-time (Fig. 1). This fact is no reason to elevate the electromagnetic field to a different ontological status. But it does explain why the field is so much harder to dispense with in the relativistic case: it is a book-keeper of events past -- and possibly even future -- rather than a summary of co-present events.

\begin{figure}[h]
	\hspace*{-0.5cm}\includegraphics[scale=0.25]{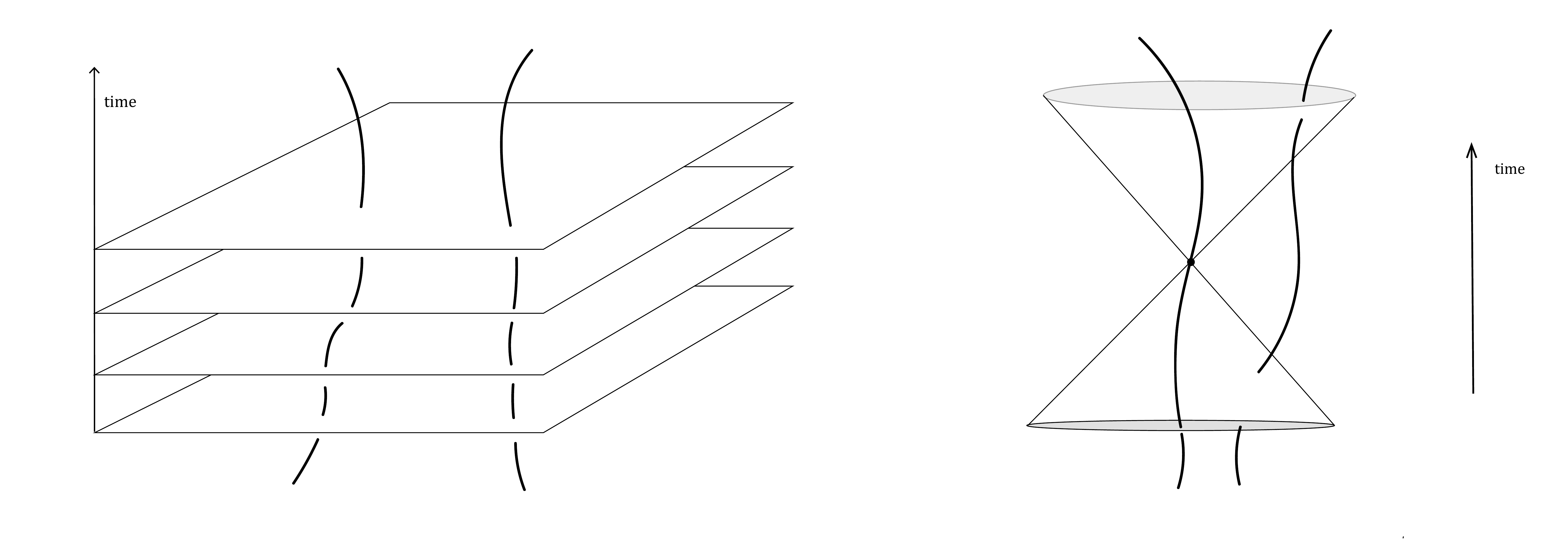}
\caption{Interactions in Newtonian (left) and Minkowski space-time (right).}
\end{figure}

 The retardation furthermore explains the appearance that electromagnetic effects \emph{propagate} at finite speed as if they had \emph{sit venia verbo} a life of their own. After all, when we look up the night sky, we see stars that have long ceased to exist. Hence, one could argue, whatever interacts with us to create the impression of a star is not the star itself (which exploded millions of years ago) but something else, viz the electromagnetic radiation, that was emitted millions of years ago and traveled all the way to earth. And while it is a natural way of speaking to say that retarded effects ``travel'' or ``propagate'', the ontological conclusion that there must be \emph{something} that actually propagates through time and space is an unwarranted application of pre-relativistic thinking.

While the retardation of electromagnetic effects is at most a practical challenge to our reductive enterprise, the possibility of free fields is a truly fundamental one since it implies that the electromagnetic field is in general \textit{not} completely determined by the charge trajectories and thus enters the particle equations of motion as an independent object. I believe that other reductive programs have not paid sufficient attention to this issue, so we will discuss it in detail in the following section.

\section{Against free fields}
Suppose we want to describe a system of charges in a space-time region $M$, where $M$ is a proper subset of Minkowski space. In general, we will have to account for the influence of charges outside of $M$ by specifying appropriate boundary conditions for the electromagnetic field (the ``incoming radiation''). These boundary conditions will necessarily contain \emph{free} or \emph{external} fields, in the sense of fields that have no sources inside of $M$ (because their sources lie outside of $M$). 

Free fields are thus essential for providing effective descriptions of subsystems. In this context, they are comparable to the external potentials / forces that we invoke in Newtonian mechanics to account for the influence of the environment otherwise ignored in the description of the system. Of course, in Newtonian mechanics, we do not believe that there are any truly external forces on the fundamental level, that is, when the relevant system is the entire universe. We rather assume that all forces experienced by a particle are ultimately attributable to interactions with other particles. (Truly external forces would in fact violate Newton's third law). As we have seen, the issue is more subtle when it comes to electrodynamics, since here, the fundamental laws already include the possibility of free fields. Nevertheless, the necessity of such fields on the fundamental level -- i.e. when $M$ is the whole of space-time -- can hardly be maintained on empirical grounds. We will never be able to determine that some observed radiation is truly source-free, coming in ``from infinity''. In fact, good scientific practice is to assume that it is \emph{not} and look for -- or simply infer -- the existence of material sources.

Empirical evidence is thus fully compatible with the assumption that the model of classical electrodynamics that best fits our universe is one in which all field degrees of freedom can be reduced to the history of charged matter. Free fields then still appear as effective devices that allow for efficient descriptions of subsystems, but there are, strictly speaking, no free fields in the universe. 

In addition to being expandable, free fields are also philosophically unsatisfying. Recall that the inhomogeneous Maxwell equation \eqref{Maxwell} allows for various different solutions. In particular,  we can consider, in addition to the retarded and advanced Li\'enard-Wiechert fields, any linear combination
\begin{equation}\label{Fieldgeneral}
(1-\lambda)\, F_{\mathrm {ret}} + \lambda\, F_{\mathrm {adv}}, \; \lambda \in [0,1].\end{equation}
Since all these fields solve \eqref{Maxwell}, the difference of any two of them is a solution of the homogeneous equation \eqref{Maxwellfree}, corresponding to a so-called free field.

Now suppose the \emph{actual} value of the total electromagnetic field at space-time point $x$ is $F(x)$. What part of this field was ``produced'' by charges and what part is due to free fields? This question actually makes no sense in the context of the field theory. If we choose the retarded inhomogeneous solution, the field attributed to the particles corresponds to $F_{\mathrm {ret}}(x)$ and we should say that the difference $F(x) - F_{\mathrm {ret}}(x)$ is the value of the ``free field''. If we choose instead the advanced Li\'enard-Wiechert solution, the field attributed to the particles is $F_{\mathrm {adv}}(x)$ and we will identify the difference $F(x) - F_{\mathrm {adv}}(x)$ as the contribution of the ``free field''. In general, choosing any linear combination \eqref{Fieldgeneral} as the so-called fundamental solution, the field attributed to charges is $(1-\lambda)  F_{\mathrm {ret}}(x) + \lambda F_{\mathrm {adv}}(x)$, while the free field corresponds to the remainder.

In the upshot, the Maxwell theory does not distinguish, in an unambiguous way, a field produced by charges from a field that exists independent of them. In particular, contrary to the common way of speaking, there are no well-defined physical quantities to which the term ``free field'' could even refer. The mathematical freedom to choose between equivalent representations of field solutions (which is a genuine feature of linear wave equations) thus leads to a \emph{physical indeterminacy} that we should be somewhat embarrassed by, if we wanted to take the electromagnetic field seriously as a causal agent. For the field theory requires us to accept that the electromagnetic field is in general not completely determined by matter, without being able to say what else there is in the world that it is determined by.

One way to avoid this conundrum is to insist that one field representation is actually distinguished. In fact, this is precisely what most textbook presentations do when they argue -- ultimately unconvincingly -- that the retarded solutions must be used for reasons of ``causality''. Emphasizing the time-symmetry of the field equation, one may instead come to the conclusion that the half advanced / half retarded representation ($\lambda = 1/2$) is physically preferred (but see \cite{Price} who comes to a different conclusion). I do not know whether other choices have ever been seriously advocated, but in principle one could insist on any other representation (say, fully advanced or 1/3 advanced + 2/3 retarded) as the ``correct'' one. 

In any case, the intuition behind such a stipulation is that there must be a precise and unambiguous way in which charges actually contribute to the electromagnetic field. An electron, let's say, must either produce retarded radiation, or retarded + advanced radiation, or... Such a move, as advocated, in particular, by \cite{Frisch2000}, was heavily criticized by \cite{Earman},  as an escape to ``suggestive but imprecise terminology'' (p. 493). Indeed, since the various representations are strictly equivalent -- they represent one and the same total field -- it would seem somewhat desperate to argue that one of them is ``more correct'' or ``more physical'' than the others.

The only way, I think, in which such intuitions can be vindicated, is if one of the field representations were distinguished \emph{by the mathematics}. And the only way, I think, in which one representation could be clearly distinguished by the mathematics, is if in that particular representation the corresponding homogeneous field was identically zero (when the charge distribution of the entire universe is taken into account). This can, in fact, be realized by assuming appropriate boundary conditions for the electromagnetic field. For instance, the Sommerfeld radiation condition, which sets the incoming fields to zero at $t=-\infty$, distinguishes the fully retarded representation.    

However, in the context of the field theory, the status of such a boundary condition remains unclear. If it holds only contingently, it cannot ground the conception that electrons produce a specific kind of radiation \emph{by necessity}. The final punchline is now that the models of classical electrodynamics in which the free fields can be eliminated, the models, that is, in which all field degrees of freedom can be reduced to the history of charged matter, are also models of a corresponding \emph{direct interaction theory}. On the fundamental level, it is thus unnecessary and unwarranted to buy into free fields as a physical possibility in the first place. Accepting the direct interaction theory instead, \emph{all} electromagnetic fields appear as mere effective devices, as secondary objects defined in terms of the particle trajectories (cf. \cite{Mundy:1989aa}, p. 46). Strictly speaking, though, there are no electromagnetic fields in the universe.

\section{The Wheeler-Feynman theory}\label{sec:WF}

 The idea to formulate classical electrodynamics by means of direct interactions goes back to \cite{Gauss1877} and was further explored by \cite{Fokker1929}, \cite{Tetrode1922} and \cite{Schwarzschild1903}. In 1945, Wheeler and Feynman showed that the time-symmetric direct interaction theory -- in which charges interact by half-retarded / half-advanced forces --  can account for all radiative phenomena captured by classical electrodynamics. Henceforth, that theory was known as \emph{Wheeler-Feynman electrodynamics}. 
 
   The Wheeler-Feynman equations of motion for an $N$ particle system read
   \begin{equation}\label{WFforce} m_k \ddot{z}^\mu_k =  \sum\limits_{j \neq k} e_k e_j \frac{1}{2} \left({}^{(j)}F^{\mu\nu}_{\mathrm {ret}} + {}^{(j)}F^{\mu\nu}_{\mathrm {adv}} \right) \dot{z}_{k,\nu}, \end{equation}
   where $ {}^{(j)}F_{\mathrm {ret}}$ and ${}^{(j)}F_{\mathrm {adv}}$ are the retarded and advanced Li\'enard-Wiechert fields of particle $j$. The interaction term can thus be understood as arising from the fundamental solution \eqref{Fieldgeneral} of Maxwell's equation with $\lambda = 1/2$, while self-interactions are excluded from the outset. 
   
   Starting from this fundamental solution of Maxwell's equation, there is, of course, an infinite number of possible direct interaction models, corresponding to different choices of $\lambda$. However, there are at least three reasons to consider the Wheeler-Feynman theory as the most serious contender:
   \begin{enumerate}[1)]
   	\item The Wheeler-Feynman theory corresponds to the manifestly \emph{time-symmetric} choice. All other choices of $\lambda$ would distinguish a particular time direction \emph{a priori}. However, since classical electrodynamics is in general understood to be fundamentally time-reversal invariant, this is a feature that one might well like to retain.  
   	
   	\item The Wheeler-Feynman theory is the only one that can be defined by a \emph{principle of least action}, which ensures a generalized form of energy-momentum conservation. The respective \emph{Fokker-Tetrode-Schwarzschild action} reads
   	\begin{equation}\label{WF action} S = \sum\limits_{i}\Bigl[-m_i\int \sqrt{\dot{z}_i^\mu \dot{z}_{i,\mu}}\,\mathrm{d}\lambda_i - \frac{1}{2}\sum\limits_{i \neq j} e_i e_j \int\int \delta\big((z_i - z_j)^2 \bigr) \dot{z}_i^\mu \dot{z}_{j,\mu} \,\mathrm{d}\lambda_i \,\mathrm{d}\lambda_j \Bigr].\end{equation}
   	\item The only obvious alternative to the time-symmetric choice would be a fully retarded direct interaction theory (corresponding to $\lambda =0$), as proposed, most prominently, by \cite{Ritz}. This theory cannot, however, explain the phenomenon of radiation damping, that is, the fact that accelerated charges lose energy-momentum (unless one admits an ad hoc modification of the equations of motion as e.g. \cite{Mundy:1989aa}). In the absence of self-fields, this radiation reaction can only come from interactions with other charges, so that the damping effect would be considerably delayed in a purely retarded theory. The time-symmetric theory is in a better position to explain radiation damping because advanced reactions to retarded actions arrive instantaneously.
   \end{enumerate}

\noindent In any case, to make good on its promises, the Wheeler-Feynman theory has to resolve two obvious challenges: (i) to account for the phenomenon of radiation damping despite the absence of a field back-reacting on accelerated charges, and (ii) to account for the radiative arrow, i.e., the fact that we observe only retarded ``radiation'' despite the fact that advanced and retarded terms figure equally into the fundamental law. \cite{Wheeler-Feynman:1945aa} address these issues in the context of their \emph{absorber theory}. They assume that an accelerated charge interacts with a large, homogeneous charge distribution surrounding it in every direction. In the absence of external disturbances, the net force from this so-called \emph{absorber} is assumed to be approximately zero. Then, Wheeler and Feynman show, in a series of three computations of increasing generality, that if the absorber particles are disturbed by retarded forces from the accelerated charge, their \emph{advanced} reaction, at sensible distances, corresponds to
\begin{equation}\label{Absorberreaction} \frac{1}{2} (F_{\mathrm {ret}} - F_{\mathrm  {adv}}). \end{equation} 
\noindent A test-particle in the vicinity of the accelerated charge will thus experience a net-effect
\begin{equation} \frac{1}{2} (F_{\mathrm {ret}} + F_{\mathrm {adv}}) +  \frac{1}{2} (F_{\mathrm {ret}} - F_{\mathrm {adv}}) = F_{\mathrm {ret}}, \end{equation}
as if the charge produced a fully retarded force. Moreover, at the location of the charge, \eqref{Absorberreaction} corresponds precisely to the radiation reaction \eqref{Schott} of the Lorentz-Dirac theory, as was already shown by \cite{Dirac}. In particular, the accelerated particle will thus experience a damping force as a result of its interaction with the absorber. 

Apart from the justification of the time-asymmetry thus introduced, the analysis could have -- and maybe should have -- ended here. Wheeler and Feynman, however, observe that the absorber response is independent of any detailed properties of the absorber, like mass or charge or exact arrangement of the particles. They therefore suggest -- having already derived the radiation reaction in a series of hands-on computations -- that it should be possible to obtain the same result from first principles and they go on to present a remarkably simple and elegant argument based on the so-called \emph{absorber condition}
\begin{equation}\label{absorbercondition}
\sum\limits_k \frac{1}{2}\left( {}^{(k)}F_{\mathrm {ret}} +\, {}^{(k)}F_{\mathrm {adv}} \right) = 0 \; \; \text{(outside the absorber)}.
\end{equation}
 This amounts to the assumption that the distribution of charges in the universe forms a \emph{complete absorber}, so that $\sum\limits_k  {}^{(k)}F_{\mathrm {ret}} = 0$ and $\sum\limits_k  {}^{(k)}F_{\mathrm {adv}} = 0$ hold separately, everywhere in empty space. 

It is quite possible that this assumption is satisfied, to good approximation, in our universe but it is only fair to point the finger at equation \eqref{absorbercondition} and ask: ``Why should we believe in that?''. Unfortunately, many commentators have thus criticized the results of Wheeler and Feynman, suggesting that their account of the radiation reaction rests on the validity of equation \eqref{absorbercondition} (e.g. \cite{Rohrlich2007}, p. 167, \cite{Earman}). This suggestion is, however, dubious because -- as shown by their first three derivations and as further emphasized in \cite{Bauer2014} -- the radiation reaction follows already from statistical assumptions that are much weaker and more robust than the infamous absorber condition. 

In any case, independent of the status of \eqref{absorbercondition}, one worry that legitimately arises is that the arguments leading to \eqref{Absorberreaction} apply also in the \emph{opposite time direction}. If we assumed that the accelerated charge interacts (by advanced forces) with an absorber in the past, the retarded absorber response would correspond to $ \frac{1}{2} (F_{\mathrm {adv}} - F_{\mathrm {ret}})$, resulting in a net-force of $F_{\mathrm {adv}}$ on nearby particles and an \emph{anti-damping} force on the accelerated charge.

Since the Wheeler-Feynman theory describes radiation damping as a many-particle phenomenon, it makes sense to look for a \emph{thermodynamic} explanation of the asymmetry. However, the argument proposed in \cite{Wheeler-Feynman:1945aa}, that seeks to rule out an advanced absorber reaction on the basis of its ``small a priori probability'' (p. 170), is generally not considered to be successful. \cite{Price} rejects it as an instance of a ``temporal double standard'' (p. 68). He points out that the seemingly conspiratorial behavior of the past absorber, whose effects would appear to converge on the charge at the precise moment it starts to accelerate, corresponds exactly to the derived response of the future absorber viewed in reverse. Hence, the probability of both processes is actually the same and we cannot rule out one over the other on purely statistical grounds. 

In a more positive review of the Wheeler-Feynman argument, \cite{Bauer2014} emphasize, also correctly, that the low probability of the absorber response corresponds to the low a priori probability of having an isolated charge, at that particular time and place, experiencing a large acceleration in the first place. Hence they suggest that the ``atypical acceleration'' experienced by the charge -- respectively the present state of the universe that allows the creation of subsystems in such atypical states of motion -- be regarded as the special (low-entropy) initial condition that gets the thermodynamic reasoning going. However, it is still unclear how this special boundary condition -- which is really more of an \emph{intermediate state} between past and future absorber rather than an \emph{initial state} -- is supposed to break the time symmetry. The central issue, emphasized in the critiques of \cite{Price}, pp. 65--73 and \cite{Arntzenius}, pp. 40--41, thus remains the following: If the retarded action from an accelerated charge in the present typically produces an advanced reaction \eqref{Absorberreaction} from the future absorber, why does the advanced action from the same charge not produce a corresponding retarded response from the past absorber?

While I do not have a definitive answer, I want to share with the reader some ideas on how this radiative asymmetry could be eventually reduced to a thermodynamic one (for a detailed and very critical discussion of other proposals, see \cite{Frisch2005}). 

First, observe that due to the time-symmetric nature of the interactions, a disturbance from the accelerated charge can affect the entire absorber at once (a series of advanced and retarded interactions can connect events at arbitrary space-like separation). Hence, while the retarded force from the accelerated charge ``propagates'' through the absorber, the absorber can simultaneously dissipate the energy and return to equilibrium due interactions between its constituting particles. Of course, the details of this process have not been described, yet. It will however suffice to assume that the dissipative process is not symmetric in time, but follows a \emph{thermodynamic arrow} along which the absorber equilibrates. We can then conclude that while experiencing the disturbance from the charge,  the ``late'' stages of the absorber are (already) in equilibrium, while the ``early'' stages of the absorber are (still) correlated with the impinging signal. This thermodynamic asymmetry would then explain the asymmetric absorber response: a retarded response from the past absorber would have to pass through the late stages (of the past absorber) that are already in equilibrium and thus opaque to outgoing radiation, while the advanced response from the future absorber are passed on through the early stages (of the future absorber) that have not yet relaxed to equilibrium (Fig. 2). We can thus conclude from the Wheeler-Feynman analysis that the radiative arrow follows the thermodynamic one.

\begin{figure}[h]
	\begin{center}
		\includegraphics[scale=0.45]{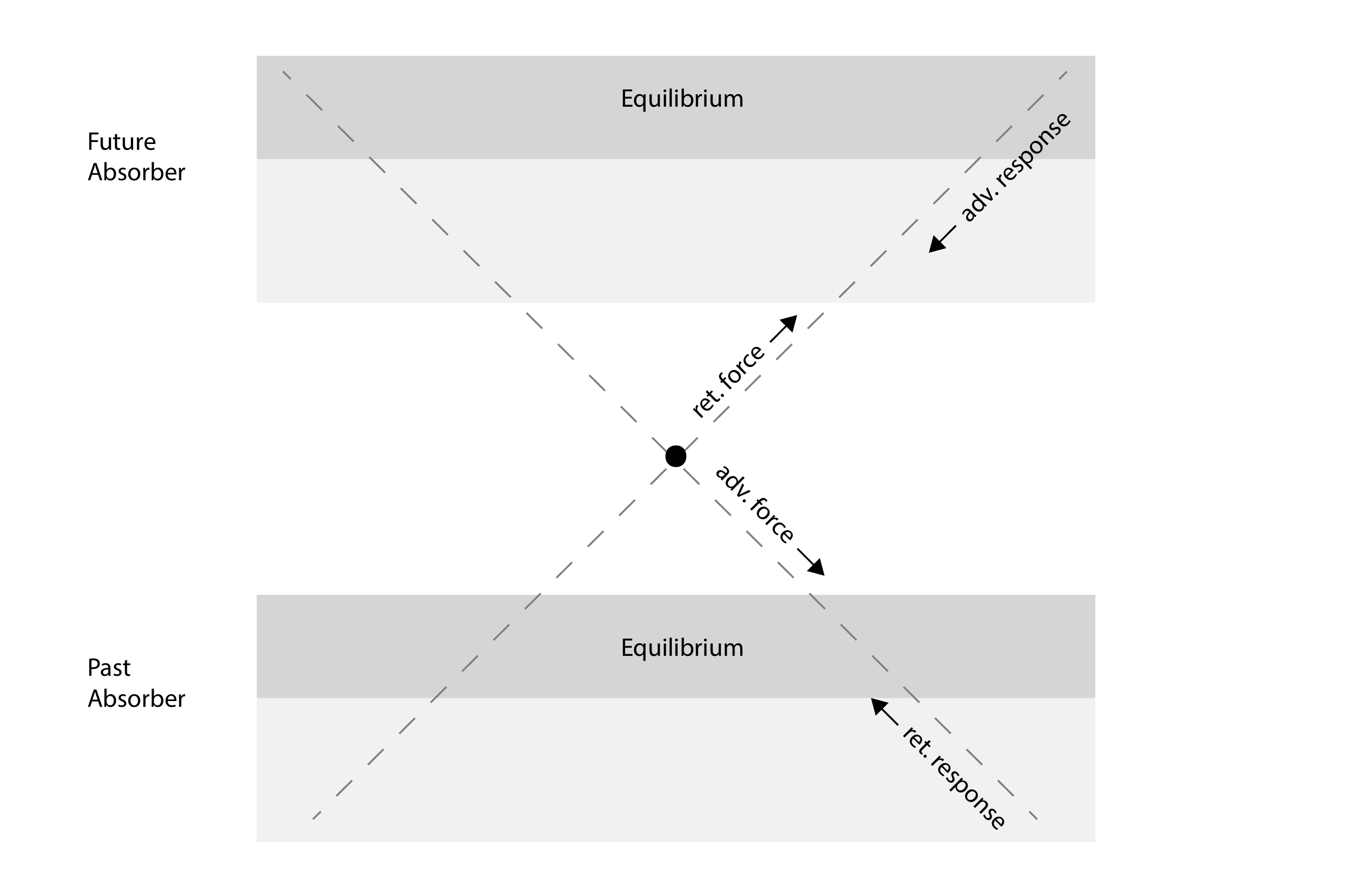}
		\caption{Absorber response to an accelerated point charge.}
	\end{center}
\end{figure}

It may also be worth pointing out that, on very large time-scales, the above picture seems to match the assumed thermodynamic evolution of our actual universe that started in a low-entropy configuration, relaxed into a state that was (roughly) in equilibrium with respect to electromagnetic interactions, has undergone structure formation under the influence of gravity and is expected to evolve into ``heath death'' in the very distant future, corresponding, once again, to a homogeneous equilibrium state.

 Regardless of the merits of the above suggestion, it seems fair to summarize that we do not yet have a conclusive account of the radiative asymmetry in Wheeler-Feynman theory (arguably, such an account is still out of reach because we know too little about the state space of the theory to formulate a precise statistical hypothesis), but that there are clear indications how the time-symmetry and the retrocausal effects manifested on the microscopic level can be reconciled with our macroscopic experience. This puts Wheeler-Feynman electrodynamics in a position that is not worse than that of the field theory, where the explanation of the radiative arrow is subject to debate, as well (for the current status of that debate, see e.g. \cite{Earman}). On the other hand, the statistical derivation of the radiation reaction is already a spectacular success that the field theory cannot match, since there, one has to rely on a highly unphysical mass renormalization to obtain analogous results. Since Wheeler-Feynman electrodynamics has no self-energy and no need for negative masses (let alone infinite ones), there are also strong indications that the theory is free of runaway solutions (\cite{Bauer:1997}). This would mean that solutions of the Lorentz-Dirac equation that are also approximate solutions of Wheeler-Feynman are automatically the good ones that do not lead to runaway behavior. In a nutshell, the direct interaction theory captures precisely the physical content of the field theory while avoiding the unphysical artifacts.

All things considered, I submit that the Wheeler-Feynman theory is by far our best candidate for a self-consistent and successful formulation of classical electrodynamics.

\section{Relativistic laws and initial value problems} \label{sec:IVP}

Here is, I think, the real reason why the Wheeler-Feynman theory is so rarely appreciated by working physicists: The Wheeler-Feynman equations of motion are not the kind of ordinary differential equations that physicists and mathematicians are trained to solve, but so-called \emph{delay differential equations}. The force acting on a particle at some space-time point $x$ depends on the trajectory of the other particles at their points of intersection with the past and future light cone of $x$ (Fig. 3); it is not determined by an \emph{instantaneous state} of the system, where ``instantaneous state'' means the configuration of the system on a space-like hypersurface that includes $x$.

\begin{figure}[h]
	\centering
	\includegraphics[scale=0.55]{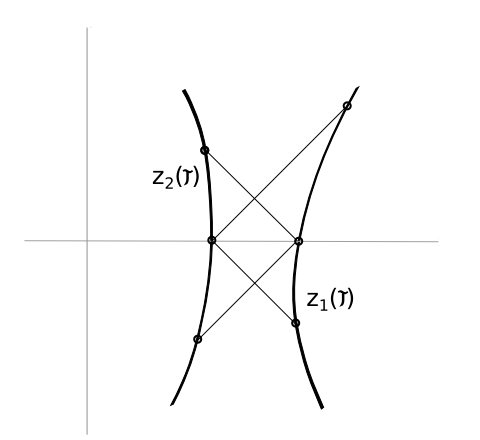}
	\caption{Direct interactions along light cones in Wheeler-Feynman electrodynamics.}
\end{figure}

As a result, the Wheeler-Feynman laws of motion cannot be brought into a Hamiltonian form, which makes the theory difficult to quantize (see \cite{Feynman:1966aa} for a nice retelling of his quest and e.g. \cite{HoyleNarlikar2} or \cite{Dirk}, ch. 8 for more recent proposals.) More generally, and more relevant to our further discussion, the Wheeler-Feynman laws are not posed as \textit{initial value problems}. As of today, it is not known what kind of boundary conditions one has to specify in order to ensure existence and uniqueness of solutions, though the best conjecture is that initial data for a Wheeler-Feynman system will comprise entire \emph{segments of trajectories}, rather than instantaneous \emph{Cauchy data} (see \cite{Bauer2013} and \cite{Deckert2016} for the current state of the solution theory). In fact, from this point of view, this is precisely the role of electromagnetic fields: to introduce additional degrees of freedom that allow the formulation of relativistic laws as initial value problems. 

Physics has grown so accustomed to initial value problems and the concept of instantaneous states that one might be tempted to take the departure from this Newtonian paradigm as a \emph{reductio ad absurdum} of the Wheeler-Feynman theory. However, it is to some extent a historical contingency that we have centuries of experience with normal differential equations, while the theory of delay equations is still underdeveloped. Moreover, one should note that for a \emph{given} distribution of charges, the effective description of a Wheeler-Feynman system corresponds to a boundary value problem in the usual Maxwell-Lorentz (or Lorentz-Dirac) theory, where fields appear as book-keeping variables in the sense discussed before. The Wheeler-Feynman theory can therefore explain the success of textbook electrodynamics and ground the familiar field formalism.

Finally and most importantly, there is really no cogent reason to expect Cauchy data in a relativistic theory. Relativistic space-time, in contrast to Newtonian space-time, does not come equipped with a foliation into instantaneous spatial geometries. The natural ontology of objects is, arguably, one of perdurant entities that are extended in time. So what, except for habit and convenience, are the reasons to expect dynamical laws that can be formulated as initial value problems? To expect, in other words, that the dynamical state of the universe is completely determined by physical data on a single space-like hypersurface? (This is not even taking into account the issue that certain \emph{general} relativistic spacetimes do not even allow a foliation into Cauchy surfaces.)

To put it differently: special relativity is usually taken to imply that \emph{all} space-like hypersurfaces (or at least hyperplanes) are equally suited for describing a complete dynamical state -- but why, in fact, should \emph{any} of them be? 

Indeed, we will now consider evidence showing that the initial value formulation is at odds with relativity, even in case of the field theory.

\subsection{The initial value problem in Maxwell's theory}\label{sec:MLIVP}
Let us take a closer look at the initial value problem in Maxwell-Lorentz electrodynamics. It will suffice to consider a single charged particle and its electromagnetic field. We shall neglect the self-interaction to avoid all problems that arise in connection with it. The theory under investigation is thus, strictly speaking, what we defined as ML-SI. 

Given a smooth field on all of space-time, the initial value problem for a test-charge is usually unproblematic. Here, we want to determine the evolution of the electromagentic field, given initial data on a space-like hypersurface $\Sigma_0$ and the particle trajectory in the future of $\Sigma_0$ (Fig. 4). For simplicity, we can assume that $\Sigma_0$ corresponds to the $t=0$ hyperplane in an appropriate coordinate system. The state of the particle at $t=0$ is $(q_0, p_0)$, where $q_0$ denotes the position and $p_0$ the momentum. It remains to specify the electromagnetic field on $\Sigma_0$. This initial field has to satisfy the \emph{Maxwell constraints} that are best expressed in terms of the electric and magnetic component as
\begin{equation}\label{Gaussgeneral} \Div E = 4 \pi \rho \;\; \; \text{and} \;\;\; \Div B = 0.\end{equation}
Here, $\rho(q) = \delta(q-q_0)$ since the charge is concentrated at the position of the particle.

Taking Maxwell's equations at face value, it would seem that \eqref{Gaussgeneral} is the only compatibility condition that the initial data for particle and field has to satisfy. From a physical point of view, there is, however, good reason to be suspicious about this. The electromagnetic field on $\Sigma_0$ must contain, in particular, the radiation emitted by the charge (in the past, if we consider the retarded representation). Due to the relativistic nature of the laws, this radiation depends on the entire (past) trajectory of the charge and is thus underdetermined by its state at one single moment in time. 

Indeed, \cite{DirkVera} recently concluded on this basis that the initial value problem in Maxwell-Lorentz electrodynamics is not well-posed. They show that only very special choices of initial data satisfying the Maxwell constraints are actually compatible, in the sense of yielding smooth solutions of the field equations, while generic initial data leads to unphysical solutions that contain discontinuities or even singularities. 

In brief, the argument can be understood as follows. Suppose we take as solution of \eqref{Gaussgeneral} the Coulomb field centered around $q_0$ (which is the default choice of most physicists). This corresponds, in fact, to the (retarded) electric field built up by a charge that has been at rest at $q_0$ since $t= -\infty$. However, the actual charge, whose field we want to describe, has in general not just been sitting there, at $q_0$, for all eternity. In particular, unless its initial momentum $p_0$ is zero, the total field obtained as a solution of Maxwell's equations will be such as if the particle experienced an infinite acceleration at $t_0$ (boosting its momentum from $0$ to $p_0$) and will thus contain a ``shock-wave'' singularity along the future light cone. Moreover, it can be shown that a higher regularity in the field requires a better and better match between the actual charge trajectory and the initial field prescribed on $\Sigma_0$. 

\begin{figure}[h]
	\begin{center}
	\includegraphics[scale=0.45]{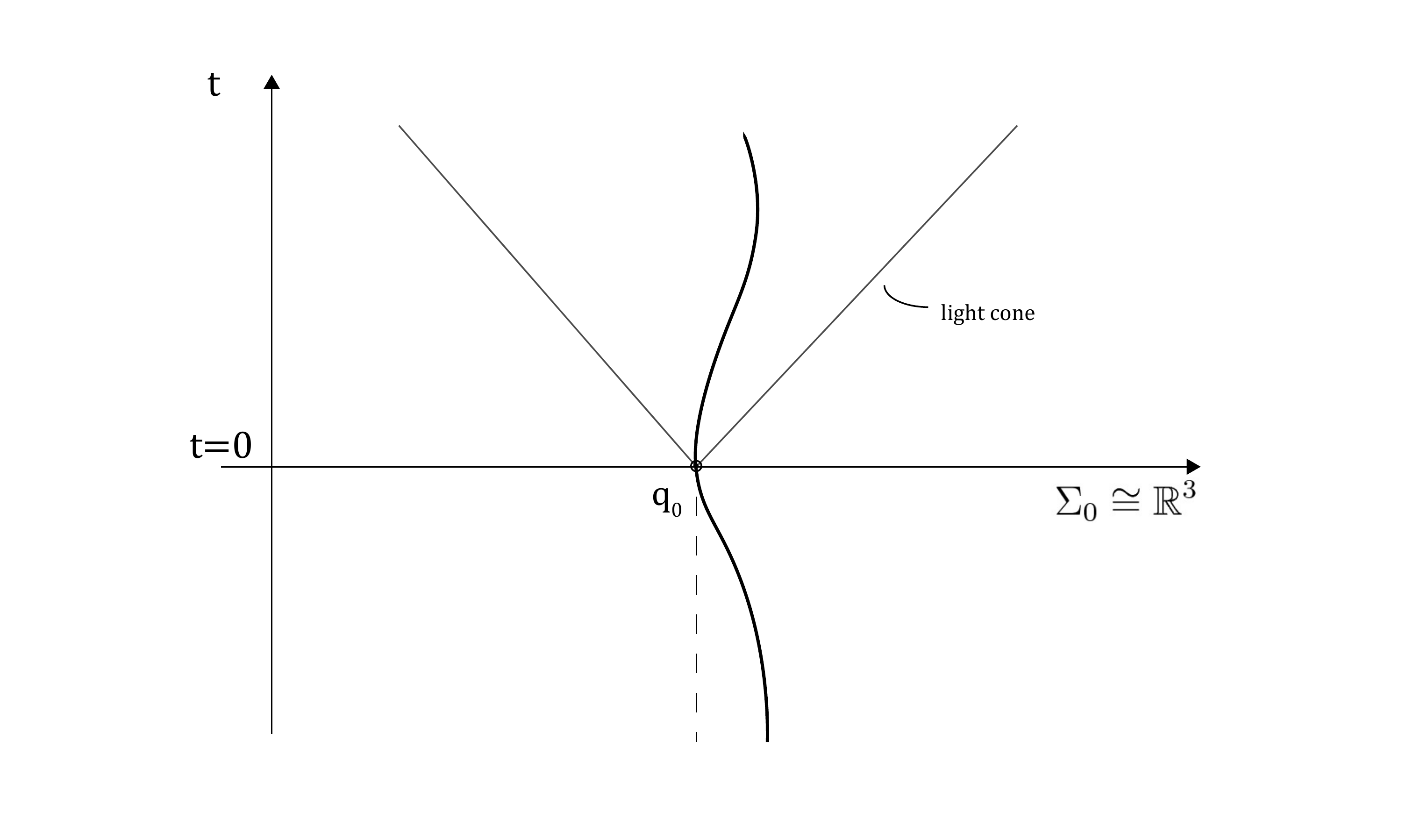}
	\caption{Initial value problem in Maxwell-Lorentz electrodynamics.}
	\end{center}
\end{figure}

\noindent We see that the initial value formulation is deceptive, as it wrongly suggests that we can forget the particle history and treat the fields as largely autonomous degrees of freedom. Indeed, Deckert and Hartenstein conclude that in order to obtain physical solutions of the Maxwell-Lorentz system, one should \emph{let go of the initial value formulation} and solve the system of delay equations that gives (the inhomogeneous part of) the field in terms of the charge trajectories and includes, in particular, the Wheeler-Feynman force law as a special case. In other words, while the correct electromagnetic field on $\Sigma_0$ would permit the formulation of an initial value problem, we do not know what the correct field is, unless we compute it from the charge trajectories in the first place (cf. \cite{Rohrlich2007}, p. 78.)


 In the upshot, the field theory allows us to trade a diachronic, spatio-temporal description in terms of particle \emph{histories} for a synchronic description in terms of an infinite number of field degrees of freedom that encode the history of charge trajectories in their spatial dependencies. The two descriptions are, however, not completely equivalent. The field formulation cannot be fundamental since typical instantaneous states allowed by the Maxwell equations describe a state of affairs that is physically impossible.\\

\noindent I promised a new narrative about field theory and the narrative that has crystallized through our analysis goes as follows: The natural form of relativistic laws is manifestly spatio-temporal, with particles interacting along light cones. Against this backdrop, the central meaning of the fields, their \emph{raison d'\^etre}, is to serve as book-keepers for the particle histories in order to save the Newtonian paradigm of instantaneous states and laws as initial value problems in a space-time geometry that does not genuinely support either. And while this works very well for practical purposes -- establishing fields as an invaluable effective device --  the failure to take the implications of relativity more seriously and the resulting dualism of particles and fields ultimately leads into various physical and metaphysical problems. In particular, the self-interaction problem, that plagues most of modern physics, must be taken as a clear indication that the field concept contains its own demise.\\

 \noindent 	\textbf{Acknowledgements} I would like to thank Michael Esfeld, Dirk-André Deckert and Mario Hubert for helpful comments and discussions. I am also grateful to Detlef Dürr, whose teachings inspired great parts of my research. This work was supported by the Cogito Foundation, grant no. 15-106-R and by a Feodor Lynen Research Fellowship of the Alexander von Humboldt Foundation. 

\newpage
\bibliographystyle{apalike}
\bibliography{againstfields}

\begin{thebibliography}{}

\bibitem[Abraham, 1903]{Abraham}
Abraham, M. (1903).
\newblock Prinzipien der {D}ynamik des {E}lektrons.
\newblock {\em Annalen der Physik}, 315(1):105--179.

\bibitem[Arntzenius, 1994]{Arntzenius}
Arntzenius, F. (1994).
\newblock Electromagnetic arrows of time.
\newblock In Horowitz, T. and Janis, A., editors, {\em Scientific Failure}.
  Rowman \& Littlefield, Maryland.

\bibitem[Bauer, 1997]{Bauer:1997}
Bauer, G. (1997).
\newblock {\em Ein Existenzsatz f{\"u}r die
  {W}heeler-{F}eynman-{E}lektrodynamik}.
\newblock Herbert Utz Verlag, M{\"u}nchen.

\bibitem[Bauer et~al., 2013]{Bauer2013}
Bauer, G., Deckert, D.-A., and D{\"u}rr, D. (2013).
\newblock On the existence of dynamics in {W}heeler--{F}eynman
  electromagnetism.
\newblock {\em Zeitschrift f{\"u}r angewandte Mathematik und Physik},
  64(4):1087--1124.

\bibitem[Bauer et~al., 2014]{Bauer2014}
Bauer, G., Deckert, D.-A., D{\"u}rr, D., and Hinrichs, G. (2014).
\newblock On irreversibility and radiation in classical electrodynamics of
  point particles.
\newblock {\em Journal of Statistical Physics}, 154(1):610--622.

\bibitem[Bell, 2004]{Bell:2004aa}
Bell, J.~S. (2004).
\newblock {\em Speakable and unspeakable in quantum mechanics}.
\newblock Cambridge: Cambridge University Press.

\bibitem[Born and Infeld, 1934]{BornInfeld}
Born, M. and Infeld, L. (1934).
\newblock Foundations of the new field theory.
\newblock {\em Proceedings of the Royal Society of London A: Mathematical,
  Physical and Engineering Sciences}, 144(852):425--451.

\bibitem[Deckert, 2010]{Dirk}
Deckert, D.-A. (2010).
\newblock {\em Electrodynamic absorber theory -- a mathematical study}.
\newblock T{\"o}nning: Der Andere Verlag.

\bibitem[Deckert and Hartenstein, 2016]{DirkVera}
Deckert, D.-A. and Hartenstein, V. (2016).
\newblock On the initial value formulation of classical electrodynamics.
\newblock {\em Preprint: arXiv:1602.0468}.

\bibitem[Deckert and Hinrichs, 2016]{Deckert2016}
Deckert, D.-A. and Hinrichs, G. (2016).
\newblock Electrodynamic two-body problem for prescribed initial data on a
  straight line.
\newblock {\em Journal of Differential Equations}, 260(9):6900--6929.

\bibitem[Dehmelt, 1988]{Dehmelt}
Dehmelt, H. (1988).
\newblock A single atomic particle forever floating at rest in free space: New
  value for electron radius.
\newblock {\em Physica Scripta}, 1988(T22):102--110.

\bibitem[Dirac, 1938]{Dirac}
Dirac, P. A.~M. (1938).
\newblock Classical theory of radiating electrons.
\newblock {\em Proceedings of the Royal Society of London A: Mathematical,
  Physical and Engineering Sciences}, 167(929):148--169.

\bibitem[D{\"u}rr et~al., 2013]{Durr:2013aa}
D{\"u}rr, D., Goldstein, S., and Zangh{\`\i}, N. (2013).
\newblock {\em Quantum physics without quantum philosophy}.
\newblock Berlin: Springer.

\bibitem[Earman, 2011]{Earman}
Earman, J. (2011).
\newblock Sharpening the electromagnetic arrow(s) of time.
\newblock In Callender, C., editor, {\em The Oxford Handbook of Philosophy of
  Time}. OUP Oxford.

\bibitem[Esfeld, 2009]{Esfeld2009}
Esfeld, M. (2009).
\newblock The modal nature of structures in ontic structural realism.
\newblock {\em International Studies in the Philosophy of Science},
  23(2):179--194.

\bibitem[Esfeld, 2014]{Esfeld:2014aa}
Esfeld, M. (2014).
\newblock Quantum {H}umeanism, or: physicalism without properties.
\newblock {\em The Philosophical Quarterly}, 64(256):453--470.

\bibitem[Esfeld et~al., 2014]{Esfeld:2014ab}
Esfeld, M., Lazarovici, D., Hubert, M., and D{\"u}rr, D. (2014).
\newblock The ontology of {B}ohmian mechanics.
\newblock {\em British Journal for the Philosophy of Science}, 65(4):773--796.

\bibitem[Feynman et~al., 1963]{Feynman1963}
Feynman, R., Leighton, R., and Sands, M. (1963).
\newblock {\em The Feynman Lectures on Physics}.
\newblock Number Vol.2 in The Feynman Lectures on Physics.
  Pearson/Addison-Wesley.

\bibitem[Feynman, 1966]{Feynman:1966aa}
Feynman, R.~P. (1966).
\newblock The development of the space-time view of quantum electrodynamics.
  {N}obel {L}ecture, {D}ecember 11, 1965.
\newblock {\em Science}, 153:699--708.
\newblock Online Version:
  \url{http://www.nobelprize.org/nobel_prizes/physics/laureates/1965/feynman-lecture.html}.

\bibitem[Field, 1985]{Field:1985aa}
Field, H.~H. (1985).
\newblock Can we dispense with space-time?
\newblock In Asquith, P.~D. and Kitcher, P., editors, {\em Proceedings of the
  1984 Biennial Meeting of the Philosophy of Science Association. Volume 2},
  pages 33--90. East Lansing: Philosophy of Science Association.

\bibitem[Fokker, 1929]{Fokker1929}
Fokker, A.~D. (1929).
\newblock Ein invarianter {V}ariationssatz f{\"u}r die {B}ewegung mehrerer
  elektrischer {M}assenteilchen.
\newblock {\em Zeitschrift f{\"u}r Physik}, 58(5):386--393.

\bibitem[Frisch, 2000]{Frisch2000}
Frisch, M. (2000).
\newblock ({D}is-)solving the puzzle of the arrow of radiation.
\newblock {\em The British Journal for the Philosophy of Science},
  51(3):381--410.

\bibitem[Frisch, 2004]{Frisch2004}
Frisch, M. (2004).
\newblock Inconsistency in classical electrodynamics.
\newblock {\em Philosophy of Science}, 71(4):525--549.

\bibitem[Frisch, 2005]{Frisch2005}
Frisch, M. (2005).
\newblock {\em Inconsistency, Asymmetry, and Non-Locality: A Philosophical
  Investigation of Classical Electrodynamics}.
\newblock Oxford University Press, New York.

\bibitem[Gau{\ss}, 1877]{Gauss1877}
Gau{\ss}, C. (1877).
\newblock A letter to {W}. {W}eber on {M}arch 19th, 1845.
\newblock In {\em Gau{\ss}: Werke}, volume~5, pages 627--–629. K{\"o}nigl.
  Gesellschaft der Wissenschaften, G{\"o}ttingen.

\bibitem[Gr{\"u}nbaum, 1976]{Grunbaum1976}
Gr{\"u}nbaum, A. (1976).
\newblock Is preacceleration of particles in dirac's electrodynamics a case of
  backward causation? {T}he myth of retrocausation in classical
  electrodynamics.
\newblock {\em Philosophy of Science}, 43(2):165--201.

\bibitem[Hoyle and Narlikar, 1969]{HoyleNarlikar2}
Hoyle, F. and Narlikar, J. (1969).
\newblock Electrodynamics of direct interparticle action. {I}. {T}he quantum
  mechanical response of the universe.
\newblock {\em Annals of Physics}, 54(2):207 -- 239.

\bibitem[Kiessling, 2012]{Kiessling2012}
Kiessling, M. K.-H. (2012).
\newblock On the motion of point defects in relativistic fields.
\newblock In Finster, F., M{\"u}ller, O., Nardmann, M., Tolksdorf, J., and
  Zeidler, E., editors, {\em Quantum Field Theory and Gravity: Conceptual and
  Mathematical Advances in the Search for a Unified Framework}, pages 299--335.
  Springer, Basel.

\bibitem[Komech and Spohn, 2000]{KomechSpohn}
Komech, A. and Spohn, H. (2000).
\newblock Long-time asymptotics for the coupled {M}axwell-{L}orentz equations.
\newblock {\em Communications in Partial Differential Equations},
  25(3-4):559--584.

\bibitem[Lange, 2002]{lange2002introduction}
Lange, M. (2002).
\newblock {\em An Introduction to the Philosophy of Physics: Locality, Fields,
  Energy, and Mass}.
\newblock Blackwell.

\bibitem[Lorentz, 1904]{Lorentz2}
Lorentz, H. (1904).
\newblock Weiterbildung der {M}axwell'schen {T}heorie: Elektronentheorie.
\newblock In {\em Enzyklop{\"a}die der Mathematischen Wissenschaften}, volume
  5, T.2, pages 145--280.

\bibitem[Maudlin, 2015]{Maudlin2015}
Maudlin, T. (2015).
\newblock The {U}niversal and the {L}ocal in quantum theory.
\newblock {\em Topoi}, 34(2):349--358.

\bibitem[Maxwell, 1865]{Maxwell}
Maxwell, J. (1865).
\newblock A dynamical theory of the electromagnetic field.
\newblock {\em J. Philosophical Transactions of the Royal Society of London},
  155:459--512.

\bibitem[Miller, 2014]{Miller:2013}
Miller, E. (2014).
\newblock Quantum {E}ntanglement, {B}ohmian {M}echanics, and {H}umean
  {S}upervenience.
\newblock {\em Australasian Journal of Philosophy}, 92(3):567--583.

\bibitem[Muller, 2007]{Muller}
Muller, F.~A. (2007).
\newblock Inconsistency in classical electrodynamics?
\newblock {\em Philosophy of Science}, 74(2):253--277.

\bibitem[Mundy, 1989]{Mundy:1989aa}
Mundy, B. (1989).
\newblock Distant action in classical electromagnetic theory.
\newblock {\em British Journal for the Philosophy of Science}, 40(1):39--68.

\bibitem[Nodvik, 1964]{Nodvik}
Nodvik, J.~S. (1964).
\newblock A covariant formulation of classical electrodynamics for charges of
  finite extension.
\newblock {\em Annals of Physics}, 28(2):225 -- 319.

\bibitem[Pietsch, 2010]{Pietsch}
Pietsch, W. (2010).
\newblock On conceptual issues in classical electrodynamics: Prospects and
  problems of an action-at-a-distance interpretation.
\newblock {\em Studies in History and Philosophy of Modern Physics}, 41:67--77.

\bibitem[Price, 1996]{Price}
Price, H. (1996).
\newblock {\em Time's Arrow and Archimedes' Point: New Directions for the
  Physics of Time.}
\newblock Oxford University Press, Oxford.

\bibitem[Ritz, 1908]{Ritz}
Ritz, W. (1908).
\newblock Recherches critiques sur l'{\'e}lectrodynamique g{\'e}n{\'e}rale.
\newblock {\em Annales de chimie et de physique}, 8(13):145--209.

\bibitem[Rohrlich, 1997]{Rohrlich1997}
Rohrlich, F. (1997).
\newblock The dynamics of a charged sphere and the electron.
\newblock {\em American Journal of Physics}, 65(11):1051--1056.

\bibitem[Rohrlich, 2007]{Rohrlich2007}
Rohrlich, F. (2007).
\newblock {\em Classical Charged Particles}.
\newblock World Scientific Publishing, Singapore, 3rd edition.

\bibitem[Schild, 1963]{Schild}
Schild, A. (1963).
\newblock Electromagnetic two-body problem.
\newblock {\em Phys. Rev.}, 131:2762--2766.

\bibitem[Schwarzschild, 1903]{Schwarzschild1903}
Schwarzschild, K. (1903).
\newblock Zur {E}lektrodynamik. ii. {D}ie elementare elektrodynamische {K}raft.
\newblock {\em Nachrichten von der {G}esellschaft der {W}issenschaften zu
  {G}{\"o}ttingen, Mathematisch-Physikalische {K}lasse}, 1903:132--141.

\bibitem[Sellars, 1963]{Sellars}
Sellars, W.~S. (1963).
\newblock Empiricism and the {P}hilosophy of {M}ind.
\newblock In {\em Science, Perception and Reality}, pages 127--196. London:
  Routledge \& Kegan Paul.

\bibitem[Spohn, 2004]{Spohn}
Spohn, H. (2004).
\newblock {\em Dynamics of Charged Particles and their Radiation Field}.
\newblock Cambridge University Press.

\bibitem[Tetrode, 1922]{Tetrode1922}
Tetrode, H. (1922).
\newblock {\"U}ber den {W}irkungszusammenhang der {W}elt. {E}ine {E}rweiterung
  der klassischen {D}ynamik.
\newblock {\em Zeitschrift f{\"u}r Physik}, 10(1):317--328.

\bibitem[Wheeler and Feynman, 1945]{Wheeler-Feynman:1945aa}
Wheeler, J.~A. and Feynman, R.~P. (1945).
\newblock Interaction with the absorber as the mechanism of radiation.
\newblock {\em Reviews of Modern Physics}, 17:157--181.

\bibitem[Wheeler and Feynman, 1949]{Wheeler-Feynman:1949}
Wheeler, J.~A. and Feynman, R.~P. (1949).
\newblock Classical electrodynamics in terms of direct interparticle action.
\newblock {\em Reviews of Modern Physics}, 21:425--433.

\end{thebibliography}

\end{document}